# Why quarks cannot be fundamental particles

C. S. Kalman

Physics Department, Concordia University
Montréal, Québec, H4B 1R6, Canada

Many reasons why quarks should be considered composite particles are found in the book *Preons* by D'Souza and Kalman. One reason not found in the book is that all the quarks except for the u quark decay. The electron and the electron neutrino do not decay. A model of fundamental particles based upon the weak charge is presented.

In January 1975, I was one among many high energy Physicists that attended the annual meeting of the APS. For the one and only time in my life, I wrote an article on an airplane trip [1]. The intent of the paper was to deal with the discovery in the past few months of the two new $\psi$ particles. Further experimental findings and the seminal theoretical work of deRujula, Georgi and Glashow [2] would ultimately convince the High Energy Physics community that these particles contained charmed quarks and that the $\psi$ particle system was analogous to positronium. To paraphrase Bjorken, protons were now to be considered as the atoms of the strong interaction. This was the end of a remarkable odyssey for me. My second paper in high-energy physics [3] published only three years earlier had been criticized by the referee for the use of the quark model to do calculations. "This paper may be termed a mathematical exercise in a physical disguise. The calculation of meson-nucleon scattering cross sections, for example is curiously at variance with experimentally established fact of the resonance dominances." Resonant dominance was an essential part of the Bootstrap Model. When that paper was published, high energy physics had still been dominated by the viewpoint of Geoffrey F. Chew and S. C. Frautschi that fundamental particles did not exist. As Gell-Mann and Ne'eman [4] put it: "The idea of the bootstrap model is that none of the 'hadrons' or strongly interacting particles is fundamental; each is merely a dynamical bound state of various combinations of hadrons including itself." Particles gave existence to each other by pulling themselves up by their own bootstraps. Hence the name of the guiding epistemology; The Bootstrap Model. At that time atomism was not favoured. In the major reference book on quarks until 1974, Kokkedee states [5]: "Of course the quark idea is ill-favoured. … The quark model should therefore at least for the moment not be taken for more than what it is namely the tentative and simplistic expression of an as yet obscure dynamics underlying the hadronic world. As such, however, the model is of great heuristic value." With the advent of the Standard Model in the late 1970's, the guiding epistemology became and still is Atomism. The essential notion of Atomism was set out in 1750 by Rudjer Boscovich: atoms contain smaller parts, which in turn contain still smaller parts, and so on down to the fundamental building blocks of matter. These fundamental particles are *indivisible* bits of matter that are ungenerated and *indestructilble*. The properties of quarks would fit the description of fundamental particles within a renewed bootstrap model (at the quark level), but are square pegs for fundamental particles as set out by Atomism. Quarks are not indestructible; some can decay into other quarks! The only place for quarks in Atomism is shown in the table:

| Structure | |
|---|---|
| Compound | Simple |
| Molecule | atom |
| Nucleus | nucleon |
| Quark | **preon** |

-

It is essential to the very atomistic underpinnings of the Standard model that quarks are composed of elementary particles: preons. Many other reasons why quarks are composed of preons are found in the book *Preons* by D'Souza and Kalman [6]. For example the standard model contains too many undetermined parameters (18) and too many particles (48).

In the table, the original atom is based upon electrical charge and the nucleon is the atom of the strong interaction. The third line of the table is likely to be based upon a third interaction, which could be the weak interaction. The usual weak interaction would then be the residual force of a super-strong force that binds the preons just as the nuclear "pionic" force is a residual force of the chromodynamic force. If so, we can put a limitation on the number of preons:

**Interaction  Charges  Entities**

| Electrical | one   | Nucleon  |
|------------|-------|----------|
| Strong     | Three | 3 quarks |
| Weak       | two   | 2 preons |

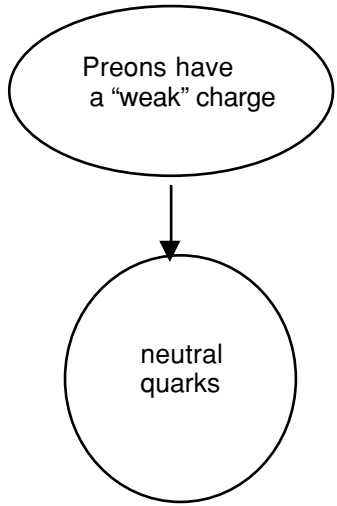

Each quark is then a $p_1 p_2$ combination. If $p_1$ and $p_2$ are both fermions, the spin of the combination is;

$1/2 \oplus 1/2 = 1, 0$  That is the quarks would have integer spins. Similarly if $p_1$ and $p_2$ are both bosons, the quarks would all have zero spin. Thus we need to combine a fermion $p_1$ with a boson $p_2$.

$$1/2 \oplus 0 = 1/2$$

This result is fortuitous because there is a severe constraint on preon models referred to as the t'Hooft anomaly matching conditions [7] These are almost impossible to meet if quarks are entirely made up of fermions. They are however easy to satisfy in models containing fermionic and scalar preons [8].

The onset of subquark structure would ultimately show up in the nucleonic x distribution. At present energies:

$$<x>_q \approx 0.3$$

At higher energies one of the quarks will be resolved into preons. Each preon contributes half of the resolved quark's momentum:

$$<x>_p \approx 0.15$$

The x distribution will thus exhibit two peaks:

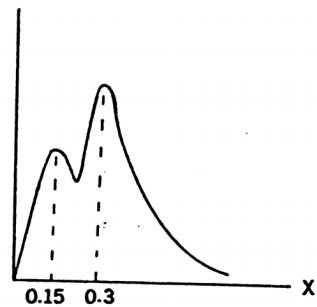

The peak centred at 0.3 corresponds to the two unresolved quarks and is thus higher than the peak corresponding to the single resolved quark.

## Conclusions

Previous discussions of preon models emphasized the desirability of preon models because of problems with the standard model.

In this paper it is emphasized that the basic notion of atomism requires the existence of preons.

It is likely that preons are based upon the weak charge. A good place to search for preons is in the nucleonic x distribution.